\documentclass[onecolumn,pra,notitlepage]{revtex4-2}
\usepackage{amsmath}
\usepackage{graphicx}
\usepackage{wrapfig}
\usepackage{enumerate}
\usepackage{float}
\usepackage{standalone}
\usepackage{lineno}
\usepackage{subcaption}

\begin{document}

\title{Optical microcavity with a pair of suspended resonant mirrors}

\author{Mikkel Kirkegaard, Trishala Mitra, Gurpreet Singh and Aur\'{e}lien Dantan}\email{dantan@phys.au.dk}

\address{Department of Physics and Astronomy, Aarhus University, DK-8000 Aarhus C, Denmark}

 %% email address is required

\begin{abstract}
We report on the realization of an optical microcavity consisting in the plane-plane arrangement of two suspended {\it resonant} mirrors possessing spectrally overlapping high-quality factor internal resonances. We first investigate its generic transmission spectra as the cavity length is varied on the basis of a simple linear Fabry-Perot model, compare them with those of broadband mirror cavities or cavities possessing a single resonant mirror, and then present an experimental realization using a pair of highly pretensioned, ultrathin silicon nitride films patterned with one-dimensional photonic crystal structures.
\end{abstract}

\date{\today}

\maketitle

%%%%%%%%%%%%%%%%%%%%%%%%%%  body  %%%%%%%%%%%%%%%%%%%%%%

\section{Introduction}

Ultracompact {\it resonant} mirrors can be realized using nanostructured thin films and engineering optical (Fano) resonances via the exploitation of interference effects between guided/localized modes in the thin film and the incoming radiation~\cite{Wang1993,Limonov2017,Quaranta2018}. Using highly pretensioned, low loss and highly reflective suspended thin films also makes it possible to take advantage of the mechanical properties of these films, which is interesting for e.g. sensing and optomechanics applications~\cite{Kemiktarak2012,Bui2012,Kemiktarak2012a,Norte2016,Reinhardt2016,Chen2017,Nair2019}.

There has recently been a strong interest in the realization of optical microcavities possessing such suspended resonant (or {\it Fano}) mirrors, i.e. ultrathin mirrors exhibiting optical internal resonances, and thereby a strongly wavelength-dependent reflectivity~\cite{Naesby2018,Cernotik2019,Fitzgerald2021,Manjeshwar2023,Mitra2024,Peralle2024,Singh2025}. Small modevolume optical microcavities with a high spectral selectivity are indeed used in a plethora of applications within optical sensing, nonlinear optics and lasers, cavity quantum electrodynamics to cavity optomechanics~\cite{Vahala2003,Aspelmeyer2014}. 

While such microcavities can be realized in a variety of integrated/in-plane architectures, they can also naturally be made in a traditional out-of-plane parallel-mirror configuration using such ultrathin suspended mirrors, which allows for preserving the mechanics of the mirrors, enhancing the strength of the light-matter interactions and ensuring efficient optical in-/out-coupling. Using high-mechanical quality membranes patterned with a one- or two-dimensional photonic crystal structure as a cavity end-mirror, optomechanical bistability and strong optomechanical interactions have recently observed in a variety of experiments~\cite{Sang2022,Xu2022,Zhou2023,Enzian2023,Manjeshwar2023,Singh2025}.

\begin{figure}[h!]
\centering
\begin{subfigure}[b]{0.31\textwidth}
\includegraphics[width=\textwidth]{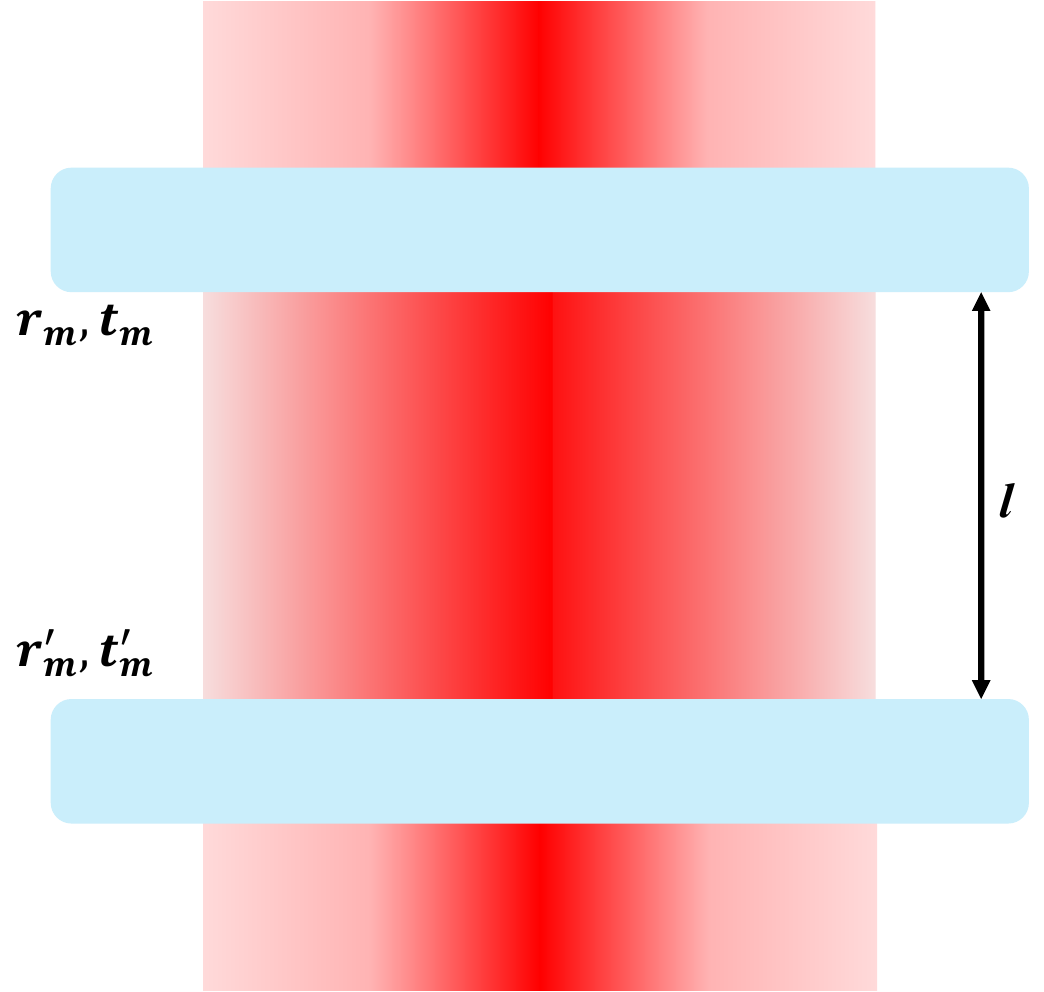}
%\caption{Single Fano cavity}
\end{subfigure}
\hspace{0.02\textwidth}
\begin{subfigure}[b]{0.31\textwidth}
\includegraphics[width=\textwidth]{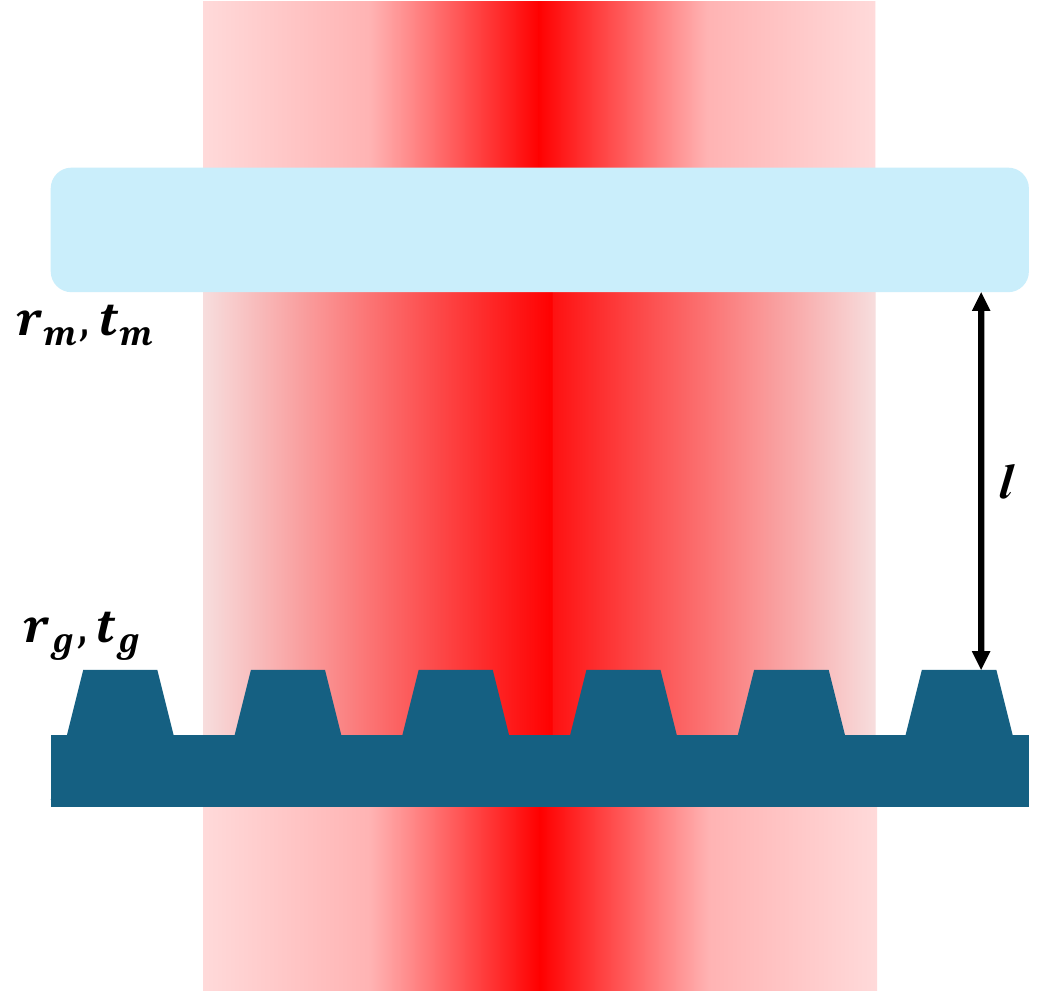}
%\caption{Single Fano cavity}
\end{subfigure}
\hspace{0.02\textwidth}
\begin{subfigure}[b]{0.31\textwidth}
\includegraphics[width=\textwidth]{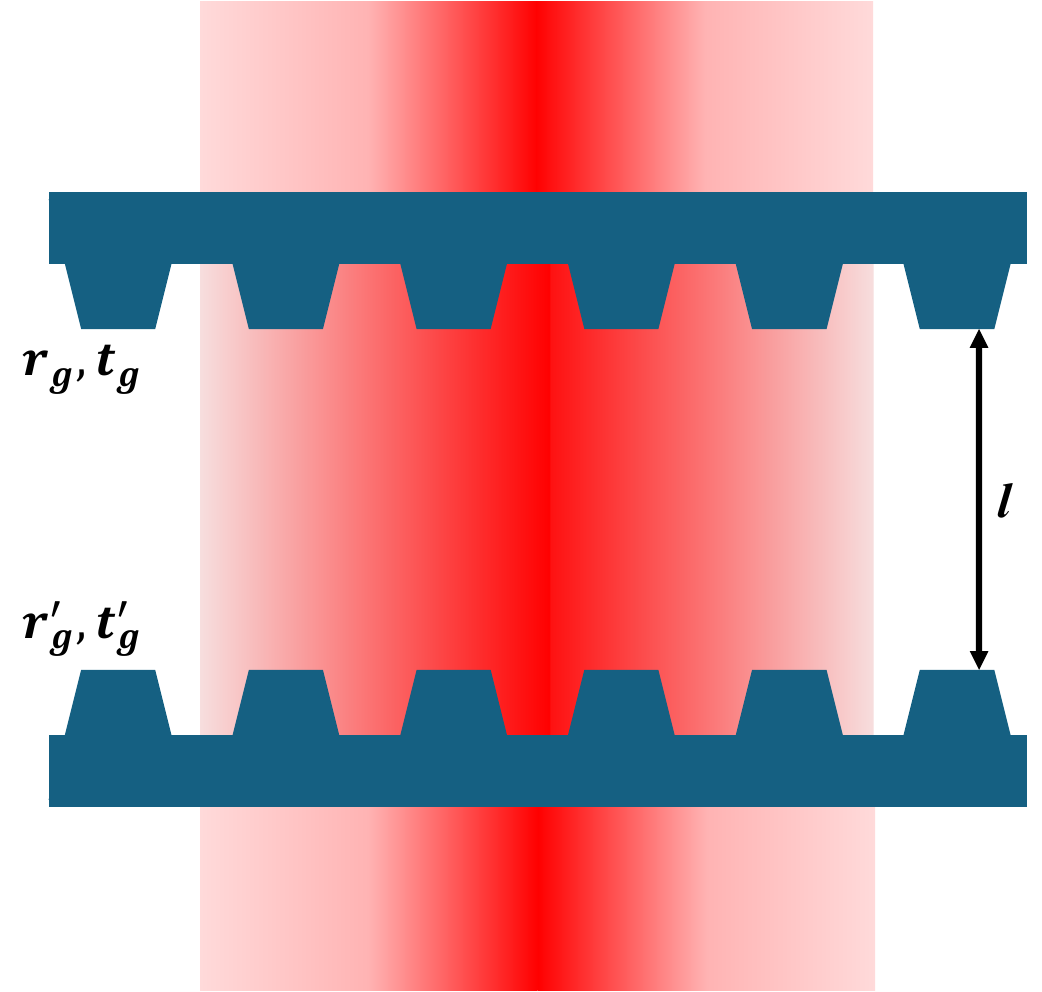}
%\caption{Double Fano cavity}
\end{subfigure}
\caption{Broadband (left), single Fano (middle) and double Fano (right) mirror cavities.}
\label{fig:broadband_single_double}
\end{figure}

In this work we report on the realization of optical microcavities consisting in the plane-plane arrangement of two suspended Fano mirrors possessing spectrally overlapping high-quality factor internal resonances. We first investigate theoretically the generic transmission properties of such cavities in comparison with those of cavities possessing either two broadband reflectivity mirrors or one broadband reflectivity mirror and a Fano mirror (Fig.~\ref{fig:broadband_single_double}). We show in particular that double Fano mirror cavities can exhibit narrower spectral linewidths than the corresponding broadband or single Fano mirror cavities at short cavity lengths, thus surpassing the linewidth reduction recently observed between broadband and single Fano mirror cavities~\cite{Mitra2024}. We then present an experimental realization using a pair of highly pretensioned, ultrathin silicon nitride films patterned with a one-dimensional photonic crystal structure.

In addition to the abovementioned applications for photonics and cavity optomechanics~\cite{Sang2022,Xu2022,Zhou2023,Enzian2023,Manjeshwar2023,Singh2025}, such double Fano mirror cavities may find applications in cavity optomechanics with multiple membrane  resonators~\cite{Xuereb2012,Xuereb2014,Nair2017,Gartner2018,Piergentili2018,Wei2019,Manjeshwar2020,Yang2020b,Sheng2021,Cao2025,Yao2025} or for gas pressure sensing~\cite{Naserbakht2019,AlSumaidae2021,Hornig2022,Salimi2024}.

The paper is organized as follows: first, a generic model for a Fano mirror possessing a high-Q resonance and for the corresponding Fano cavities is introduced in Sec.~\ref{sec:theory}. A comparison of their transmission spectra as the cavity length is varied and a discussion of the effect of non-spectrally matched Fano mirrors are also given. In Sec.~\ref{sec:experiment} the experimental realization with Fano mirrors consisting in silicon nitride thin films patterned with subwavelength gratings is discussed; first, the experimental setup and cavity alignment procedure are introduced before experimental cavity transmission spectra and cavity resonance linewidths are discussed. A brief conclusion is given in Sec.~\ref{sec:conclusion}.

%%%%%%%%%%%%%%%%%%%%%%%%%%%%%%%%%%%%%%%%%%%%%%%%%%%%

\section{Fano cavities with resonant mirrors}
\label{sec:theory}

\subsection{Resonant Fano mirror model}
\label{sec:grating_model}

We first consider a resonant Fano mirror consisting of a subwavelength grating, surrounded by air on both sides and illuminated at normal incidence by a linearly polarized monochromatic beam, with a wavelength close to a guided-mode resonance wavelength. We describe the normal incidence transmission and reflection amplitude coefficients of the grating around using the model of~\cite{Mitra2024}, which is a generalization of the model of~\cite{Fan2003,Bykov2015} to a lossy structure taking into account potential losses due to finite grating and beam size~\cite{ToftVandborg2021,Mitra2024}.
\begin{equation}
t_g=t_d+\frac{a}{k-k_1+i\gamma}\hspace{0.5cm}\textrm{and}\hspace{0.5cm} r_g=r_d+\frac{b}{k-k_1+i\gamma},
\end{equation}
where $t_d$ and $r_d$ are the direct transmission and reflection coefficients of the slab far from resonance (satisfying $|t_d|^2+|r_d|^2=1$), $k=2\pi/\lambda$ is the incident light wavenumber, $k_1=2\pi/\lambda_1$ is the guided-mode resonant wavenumber and $\gamma$ determines the width of the guided-mode resonance. $a$ and $b$ are complex coefficients describing the interference between the direct transmitted/reflected waves and the guided-mode and taking into account collimation and finite-size losses~\cite{ToftVandborg2021,Mitra2024}. Following Ref.~\cite{Mitra2024} we rewite the grating transmission coefficient as
\begin{equation}
t_g=t_d\frac{k-k_0+i\beta}{k-k_1+i\gamma},
\end{equation}
where the resonant wavenumber $k_0$ and the constant $\beta$ are defined from $a=t_d(k_1-k_0+i\beta-i\gamma)$. Assuming resonant guided-mode losses, the energy balance relation for the grating can be expressed as
\begin{equation}
|t_g|^2+|r_g|^2+\frac{c^2}{(k-k_1)^2+\gamma^2}=1,
\end{equation}
where the constant $c$ is related to the resonant loss level $L$ by $c^2=L((k_0-k_1)^2+\gamma^2)$. Assuming for simplicity $t_d$ and $r_d$ real, the real and imaginary part of the reflectivity coefficient, $x_b$ and $y_b$ can be found by solving
\begin{align}
\label{eq:xb_general} & t_dx_a+r_dx_b=0,\\
\label{eq:yb_general} & x_a^2+y_a^2+x_b^2+y_b^2+c^2+2t_d\gamma y_a+2r_d\gamma y_b=0,
\end{align}
where $x_a$ and $y_a$ are the real and imaginary part of $a$, respectively.

\begin{figure}
\centering\includegraphics[width=0.6\textwidth]{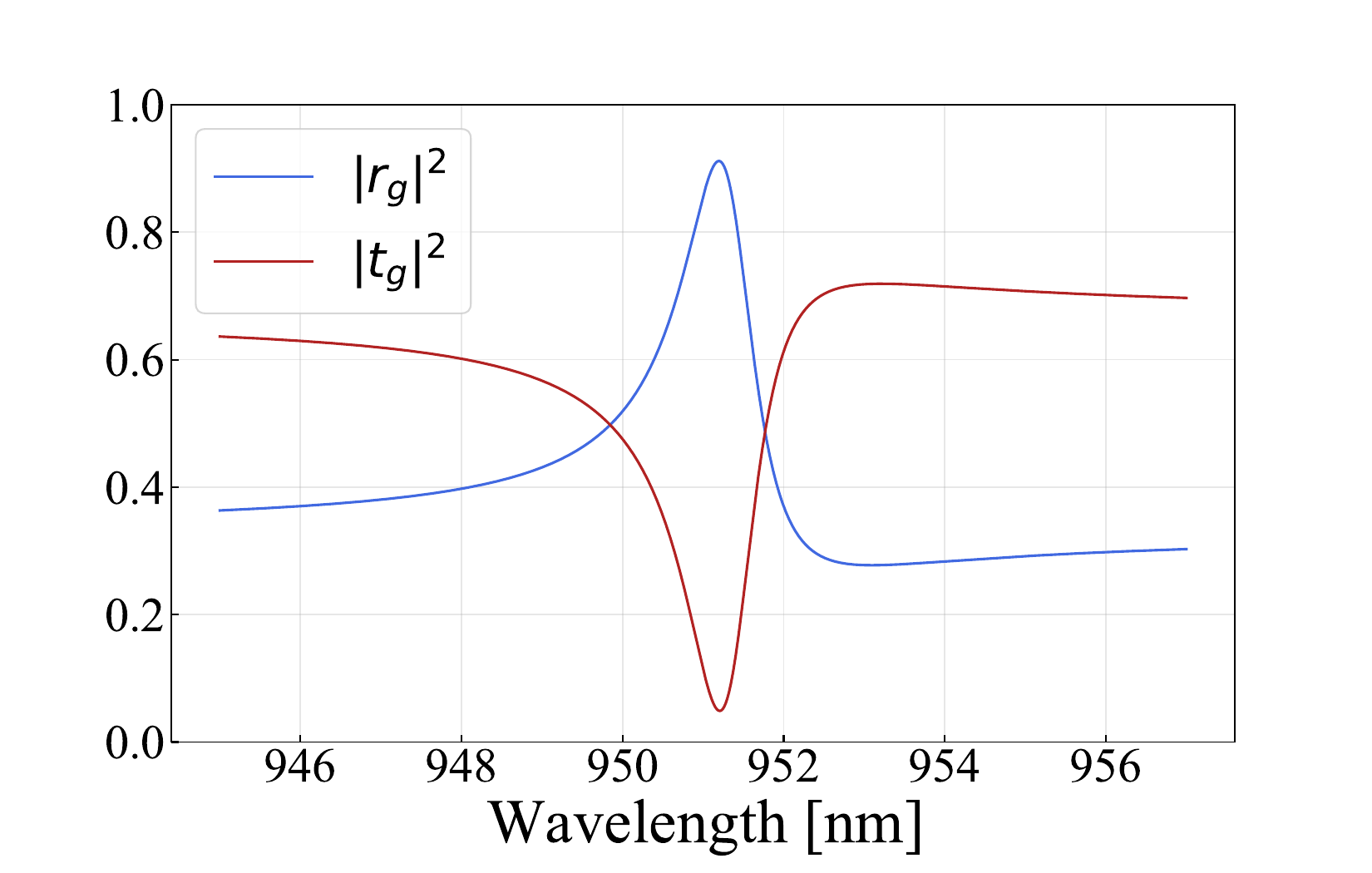}
\caption{Transmission (red) and reflectivity (blue) spectra of the lossy resonant grating described in the text, which exhibits a high-reflectivity resonance around 951.2 nm with a quality factor of $\sim900$ and a peak reflectivity of 91.2\%.}
\label{fig:spectra_grating_sim}
\end{figure}

Figure~\ref{fig:spectra_grating_sim} shows an example of transmission and reflectivity spectra of a generic subwavelength grating with parameters close to those used in the experiments. The model parameters used are $t_d=0.818$, $\lambda_0=951.23$ nm, $\lambda_1=951.36$ nm, $\gamma_\lambda=0.527$ nm, $\beta=1.03\times 10^{-6}$ nm$^{-1}$, corresponding to a minimum transmission and loss levels of 4.9\% and 4.0\%, respectively, and resulting in a peak reflectivity of 91.2\% at $\lambda_0$. The quality factor of the Fano resonance is $\sim 900$. We will use these parameters as a basis for the discussion of the generic behavior of single- and double-Fano mirror cavities based on such resonant mirrors in the following section.

\subsection{Single- {\it vs} double-Fano mirror cavities}
\label{sec:single_double}

In order to model the transmission spectrum of the various cavities shown in Fig.~\ref{fig:broadband_single_double} we use the planar Fabry-Perot transmission function
\begin{align}
\mathcal{T}_\textrm{cav}=\left|\frac{tt'e^{i\phi}}{1-rr' e^{2i\phi}}\right|^2
\label{eq:Tcav}
\end{align}
where $t, t'$ and $r, r'$ denote the incoupling and outcoupling mirror transmission and reflection coefficients at normal incidence, and $\phi=kl$, where $l$ is the cavity length.\\

For a \textit{symmetric broadband mirror cavity} with $t=t'=t_m$ and $r=r'=r_m$ independent of wavelength in the range considered around $\lambda_0$, the cavity transmission spectrum exhibits Lorentzian-shaped resonances separated by a Free Spectral Range (FSR) given by $\lambda_0/2l$ and with a HWHM
\begin{align}
\delta\lambda_c=\frac{\lambda_0^2}{8\pi l}\mathcal{L},
\label{eq:deltalambda_b}
\end{align}
where $\mathcal{L}=2(1-R_m)$ and $R_m=|r_m|^2$ represents the (intensity) reflectivity level of the broadband mirror at resonance.\\

For a \textit{single Fano mirror cavity} for which $(t,r)=(t_g,r_g)$ and $(t',r')=(t_m,r_m)$, the cavity transmission spectrum is similar to that of the corresponding broadband mirror cavity as long as the cavity FSR is much smaller than the Fano mirror resonance width. However, as the cavity length is reduced and the FSR increases, fewer cavity modes can be found in the region where the Fano mirror possesses a high reflectivity. For a cavity  resonance close to the grating zero-transmission resonance $\lambda_0$ the cavity transmission is well-approximated by
\begin{align}
\mathcal{T}_\textrm{cav}\simeq\frac{A}{1+\left(\frac{\Delta}{1-\nu\Delta}\right)^2}+B,
\label{eq:cavitylinewidth}
\end{align}
where $A$ and $B$ are constants, $\Delta=(\lambda-\lambda_c)/\delta\lambda$ is the wavelength detuning (normalized by the HWHM $\delta\lambda$) from the cavity resonance $\lambda_c$, and $\nu$ is a constant determining the degree of asymmetry of the Fano transmission profile. One can show that, when $\lambda_c\simeq\lambda_0$, the cavity HWHM $\delta\lambda$ is approximately given by
\begin{align}
\delta\lambda\simeq\frac{1}{\frac{1}{\delta\lambda_c}+\frac{1}{\delta\lambda_g}},
\label{eq:deltalambda}
\end{align}
where $\delta\lambda_c$ is the HWHM of the corresponding broadband cavity with the same length and total cavity losses $\mathcal{L}=T_m+T_g+L=2-R_m-R_g$, $T_g=|t_g(\lambda_0)|^2$ and $T_m=|t_m|^2$ represent the intensity transmission levels of the grating and the mirror at resonance, and
\begin{align}
\delta\lambda_g=\frac{\gamma_\lambda}{2(1-r_d)}\mathcal{L}
\label{eq:deltalambda_f}
\end{align}
is the HWHM of the Fano mirror cavity in the Fano regime, where $\gamma_\lambda=(\lambda_1^2/2\pi)\gamma$ represents the grating resonance HWHM in wavelength. Equation~(\ref{eq:deltalambda}) shows clearly that, for long cavity lengths, the Fano mirror cavity linewidth is given by that of the corresponding broadband mirror cavity, whereas, for short cavity lengths, it is given by Eq.~(\ref{eq:deltalambda_f}), which shows that is becomes independent of the length and solely determined by the Fano mirror resonance width and the total cavity losses.\\

\begin{figure}[h]
\begin{subfigure}{0.49\columnwidth}
\includegraphics[width=\columnwidth]{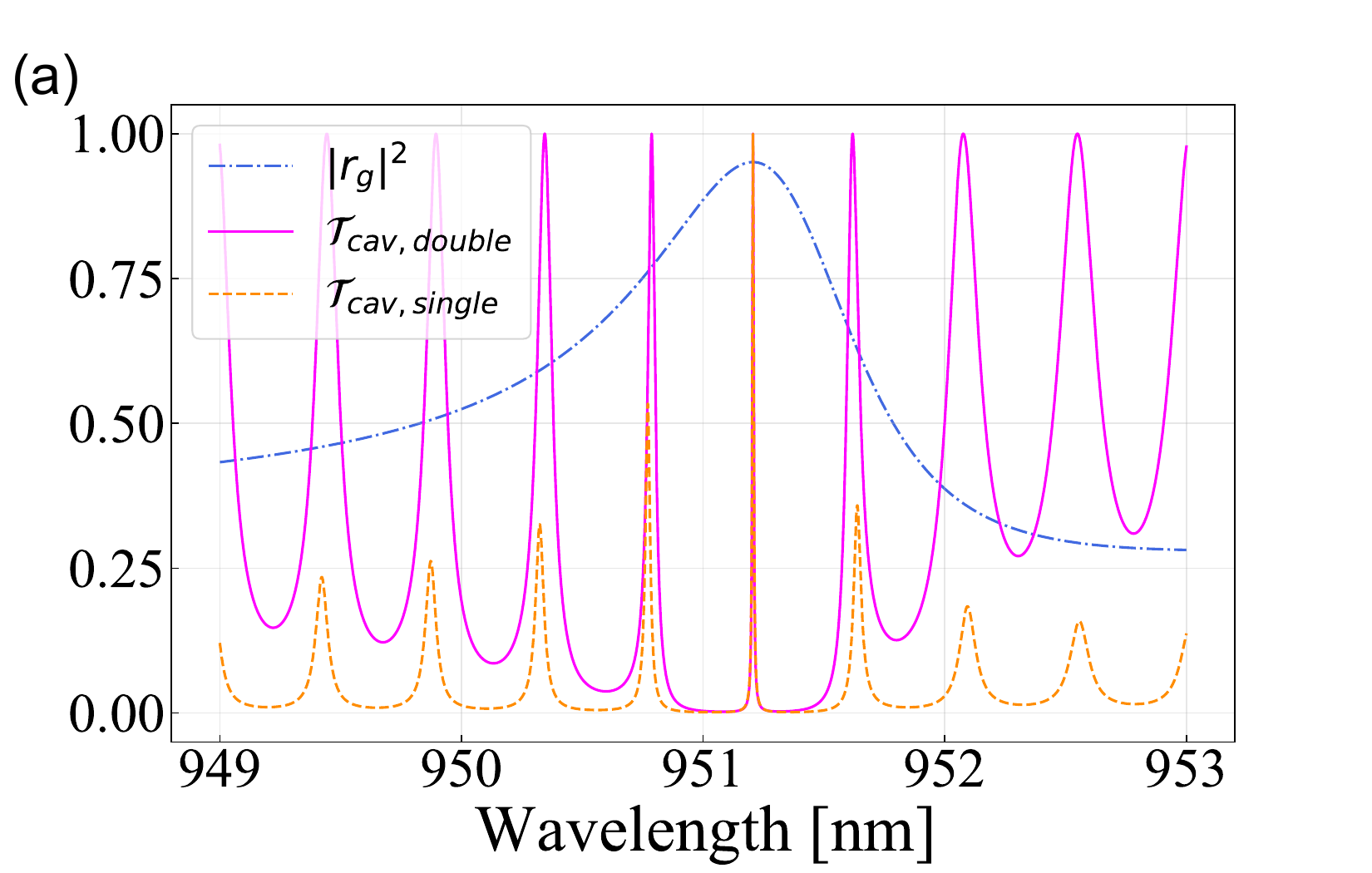}
\end{subfigure}
\begin{subfigure}{0.49\columnwidth}
\includegraphics[width=\columnwidth]{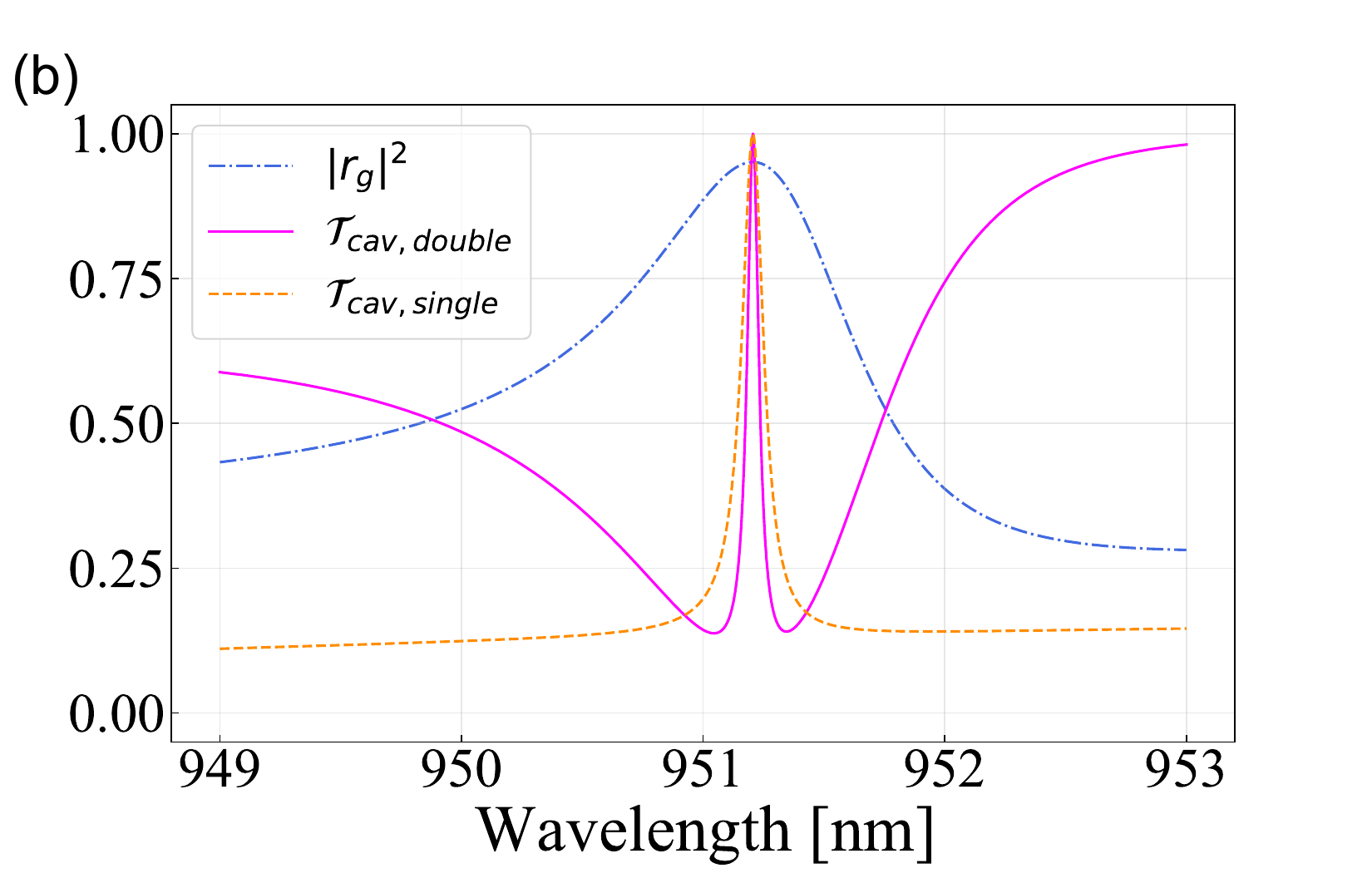}
\end{subfigure}
\begin{subfigure}{0.49\columnwidth}
\includegraphics[width=\columnwidth]{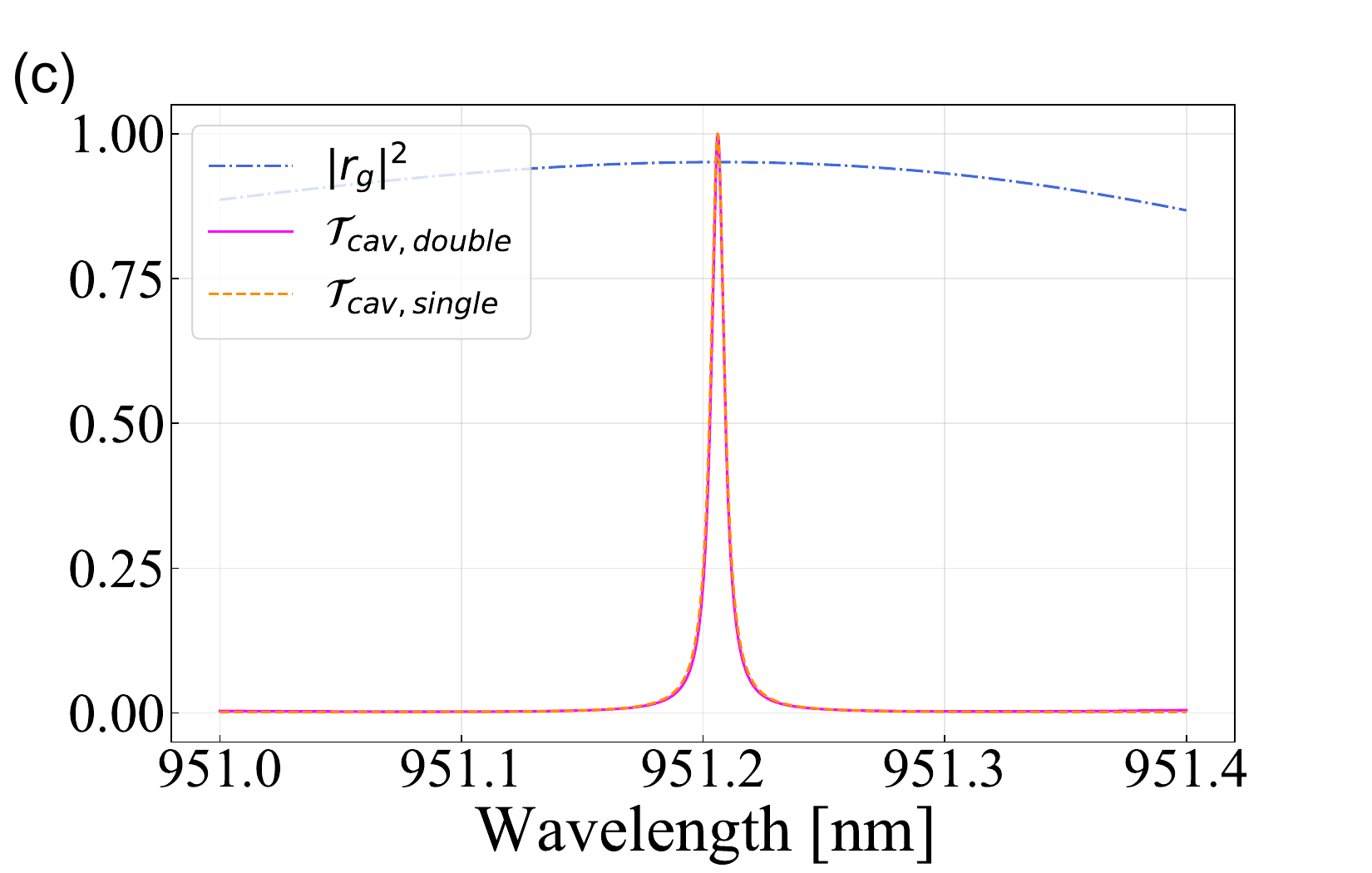}
\end{subfigure}
\begin{subfigure}{0.49\columnwidth}
\includegraphics[width=\columnwidth]{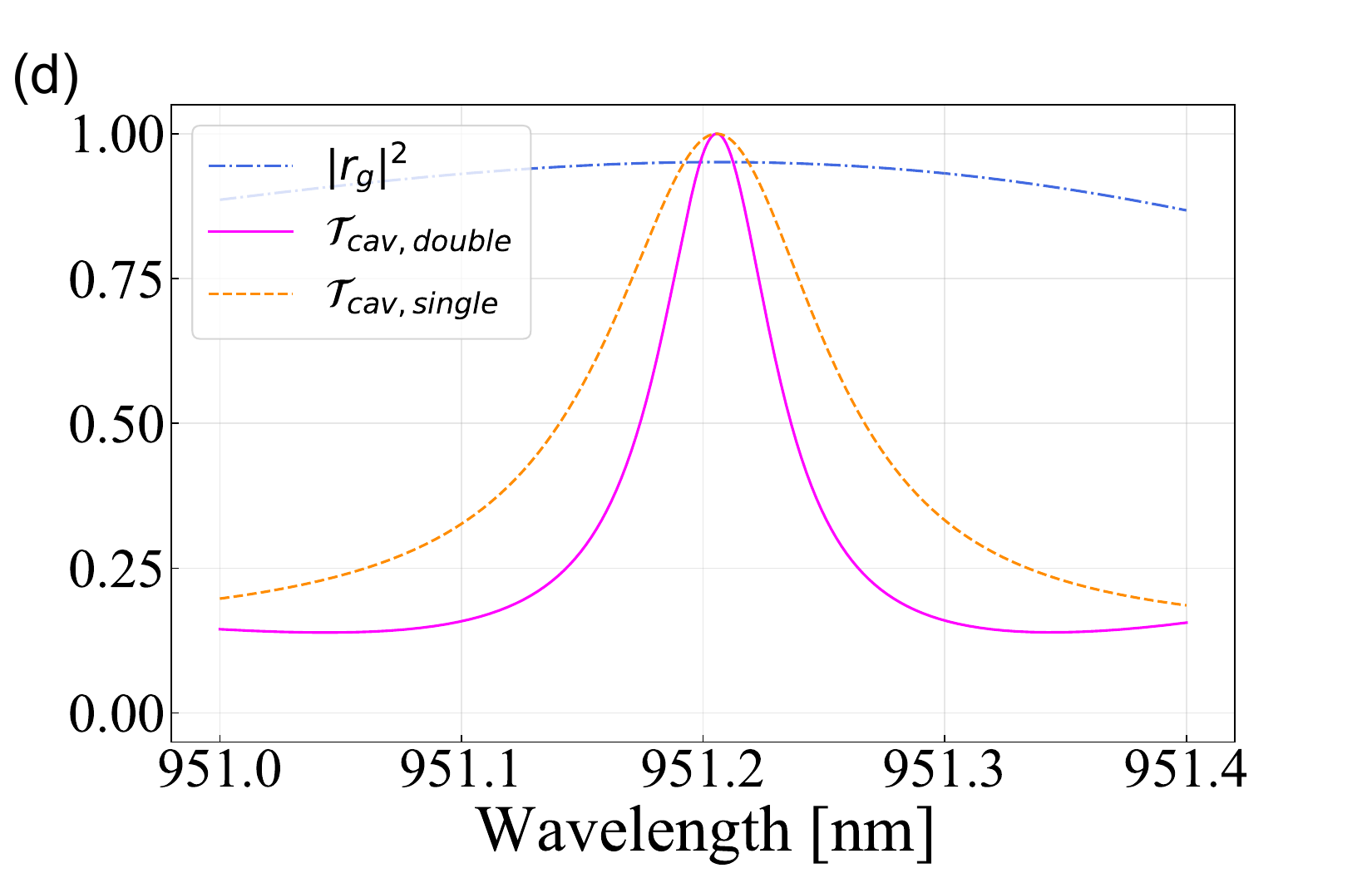}
\end{subfigure}
\caption{Transmission spectra of single Fano (orange) and double Fano (magenta) mirror cavities for lengths $l=1$ mm (a) and $l=5.2$ $\mu$m (b). The Fano mirror, whose reflectivity is shown as the dashed blue line, is that of Fig.~\ref{fig:spectra_grating_sim}. For the single Fano mirror cavity, the reflectivity of the broadband mirror is chosen to be equal to the Fano mirror peak reflectivity (91.2\%). (c) and (d): close ups of (a) and (b) around the Fano resonance.}
\label{fig:regimes}
\end{figure}

\begin{figure}
\centering
\includegraphics[width=0.6\textwidth]{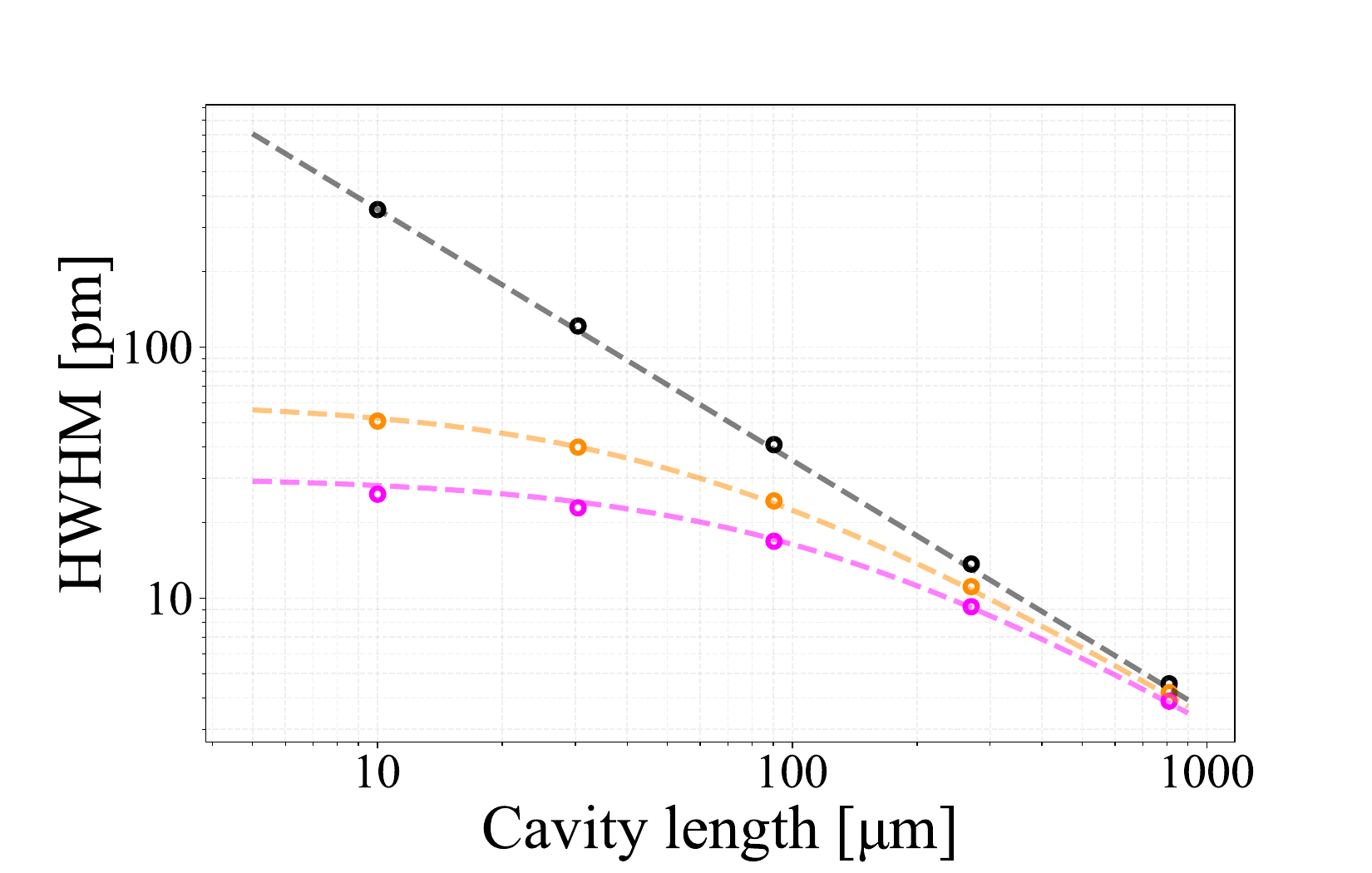}
\caption{Simulated cavity linewidth (HWHM) as a function of its length for a broadband mirror cavity (black circles), single Fano mirror cavity (orange circles) and double Fano mirror cavity (magenta circles). In all cases, the total cavity losses at the Fano resonance wavelength are 17.6\%. The points show the results of fits with Eq.~(\ref{eq:cavitylinewidth}) of the numerically simulated spectra using Eq.~(\ref{eq:Tcav}), while the black, orange and magenta dashed lines show the results of the analytical predictions of Eqs.~(\ref{eq:deltalambda_b}), (\ref{eq:deltalambda}) and (\ref{eq:deltalambda2}), respectively.}
\label{fig:length}
\end{figure}

For a \textit{symmetric double Fano mirror cavity} for which $t=t'=t_g$ and $r=r'=r_g$, a similar behavior is also observed when the cavity length is chosen so that the cavity resonant wavelength matches that of the Fano mirrors. However, away from the Fano mirror resonance, the double Fano mirror cavity transmission spectrum is similar to that of a low finesse Fabry Perot resonator, since the Fano mirror reflectivity is given by its (typically far from unity) off-resonant level $|r_d|^2$. Moreover, as illustrated on Fig.~\ref{fig:regimes}(d), the linewidth of the double Fano mirror cavity in the Fano regime is about twice as narrow as that of the corresponding single Fano mirror cavity with the same total cavity losses. Indeed, for a double Fano mirror cavity with identical Fano mirrors, we expect the cavity linewidth around the Fano resonance wavelength to be approximately given by
\begin{equation}
\delta\lambda\simeq\frac{1}{\frac{1}{\delta\lambda_c}+\frac{1}{2\delta\lambda_g}},
\label{eq:deltalambda2}
\end{equation}
where $\delta\lambda_c$ and $\delta\lambda_g$ are still given by (\ref{eq:deltalambda_b}) and (\ref{eq:deltalambda_f}), respectively, but with $\mathcal{L}=2T_g+2L=2(1-R_g)$. This is corroborated by Fig.~\ref{fig:length}, which shows the results of a systematic simulation of the linewidths of the three types of cavities as their length is varied and for same total cavity losses on resonance.

\subsection{Non-overlapping Fano resonances}

So far, we have discussed the case of double Fano mirror cavities consisting of a pair of {\it identical} Fano mirrors. However, a practically relevant scenario, especially when Fano mirrors possessing high optical quality factor Fano resonances are at play, is that of Fano mirrors with near, but not perfectly degenerate Fano resonances. To simplify the discussion and to relate to the experimental results presented in the next section, let us consider the case of two Fano mirrors which differ only by the value of their Fano resonance wavelength $\lambda_0$. We thus consider a double Fano mirror cavity consisting in the plane-plane arrangement of the Fano mirror whose model parameters ($t_d$, $\lambda_0$, $\lambda_1$, $\gamma_\lambda$, $\beta$, $L$) are those of Sec.~\ref{sec:grating_model} and a Fano mirror with the same parameters, except that $\lambda_0'=\lambda_0+\Delta$ and $\lambda_1'=\lambda_1+\Delta$, to account for a potential spectral offset $\Delta$ between the two Fano resonance wavelengths.

\begin{figure}
\centering
\includegraphics[width=\columnwidth]{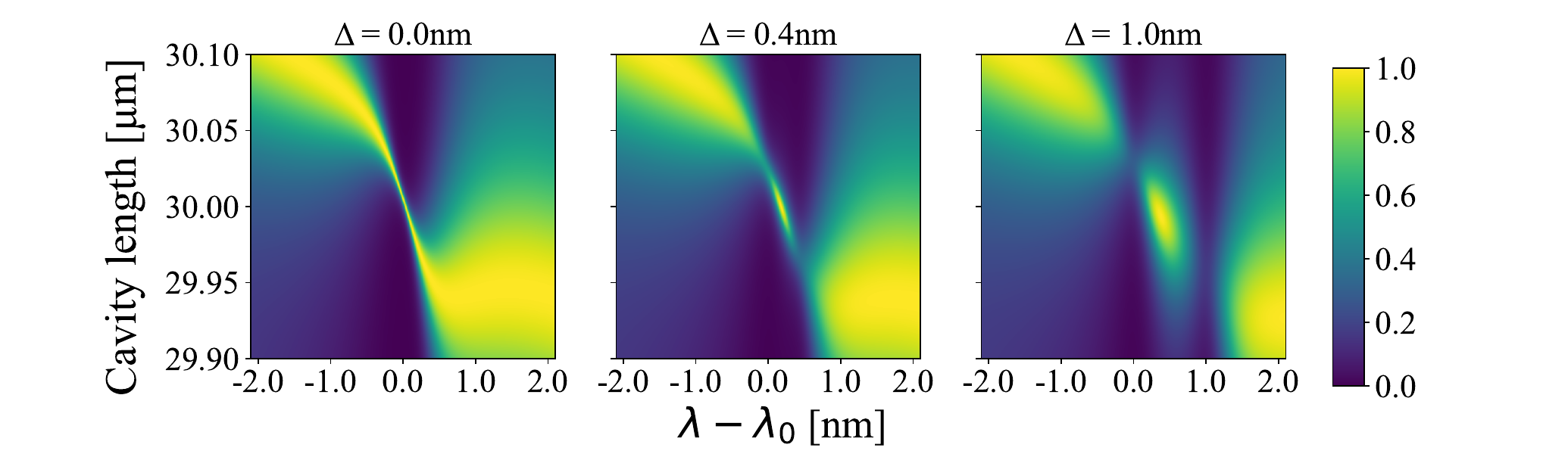}
\caption{Double Fano mirror cavity transmission as a function of cavity length and wavelength for spectrally degenerate ($\Delta=0$, left) and spectrally non-degenerate ($\Delta=0.4$ nm (middle); $\Delta=1$ nm, right) Fano mirrors.}
\label{fig:cmap}
\end{figure}

Figure~\ref{fig:cmap} shows the cavity transmission as a function of wavelength and cavity length for spectrally degenerate and non-degenerate Fano mirrors. As expected, Fig.~\ref{fig:cmap}(a) shows that, for $\Delta=0$, the narrowest linewidth is obtained when the cavity length is chosen so that the cavity resonance wavelength corresponds to the Fano mirrors resonance wavelength. When the Fano mirrors are (slightly) spectrally non-degenerate (Figs.~\ref{fig:cmap}(b) and (c)), the narrowest linewidths are obtained when the cavity length is chosen so the cavity resonant wavelength is approximately the average of both Fano resonant wavelengths ($\lambda_c\simeq(\lambda_0+\lambda_0')/2$). The narrowest cavity linewidth is also observed to rapidly broaden as $\Delta$ is increased, stressing the necessity of having highly spectrally overlapping Fano mirrors in order to achieve a significant linewidth narrowing in the Fano regime with a double Fano mirror cavity.

%%%%%%%%%%%%%%%%%%%%%%%%%%%%%%%%%%%%%%%%%%%%%%%%%%%%

\section{Experimental methods and results}
\label{sec:experiment}

\subsection{Subwavelength grating Fano mirrors}

\begin{figure}[h]
\centering
\includegraphics[width=0.55\columnwidth]{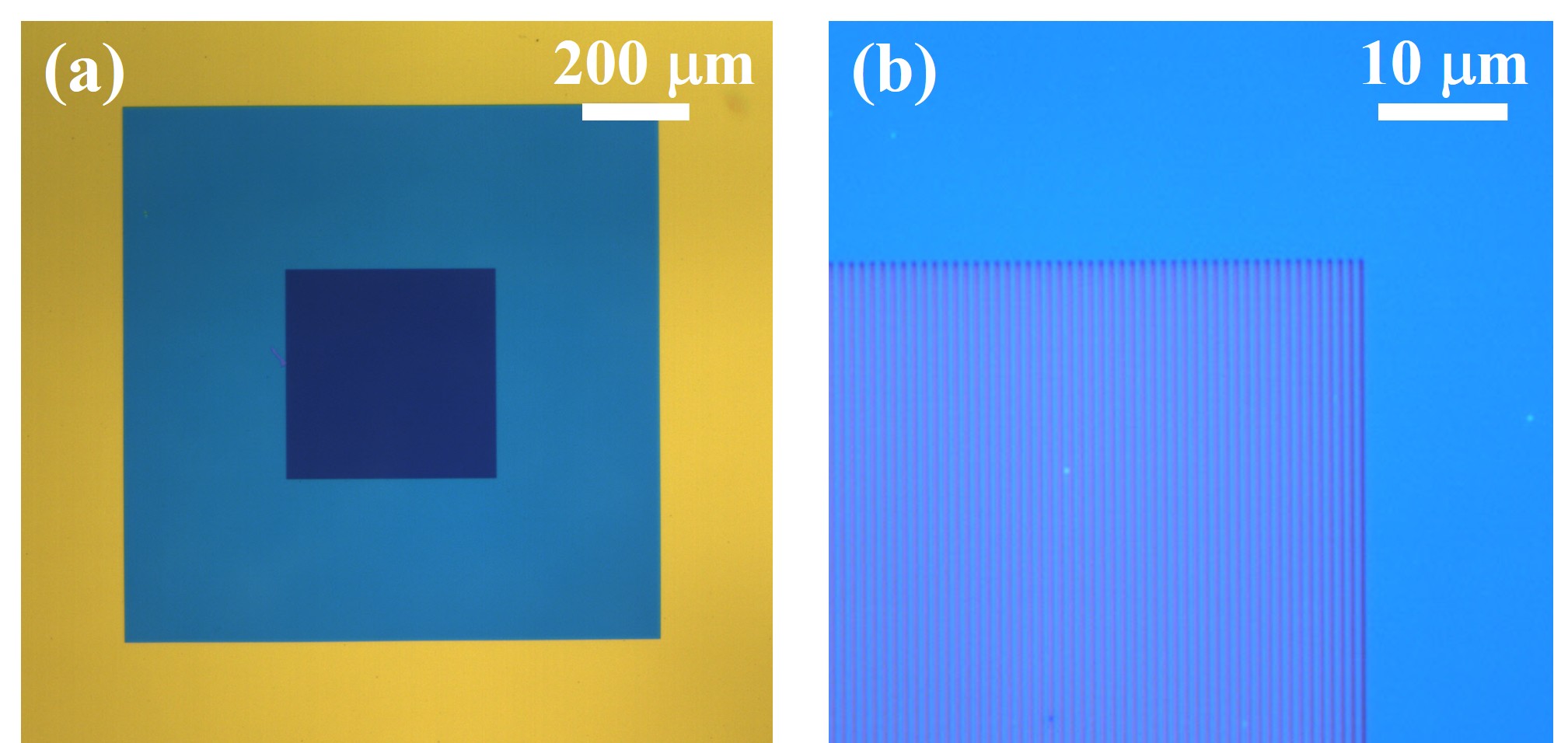}\\
\includegraphics[width=0.6\columnwidth]{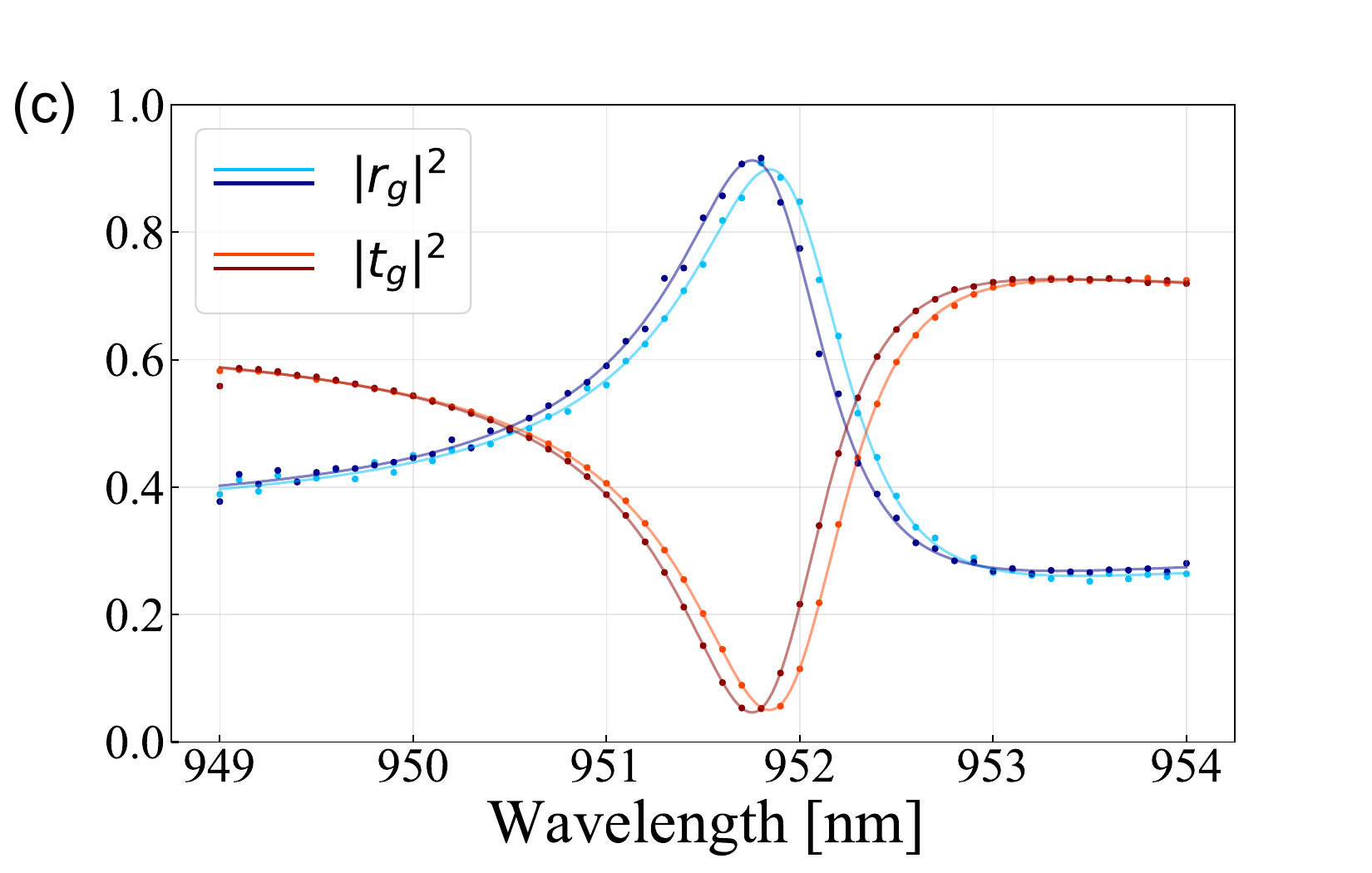}
\caption{(a) Photograph of suspended SiN membrane (blue) patterned with a subwavelength grating (dark blue). (b) Zoom-in on the edge of the patterned area. (c) Measured transmission and reflectivity spectra of Fano mirrors $A$ (orange and cyan) and $B$ (red and blue). The lines show the results of fits with the theoretical model.}
\label{fig:gratings}
\end{figure}

In this section and the next we report on experiments carried out with two similar Fano mirrors, in the form of highly pretensioned, ultrathin silicon nitride films patterned with one-dimensional photonic crystal structures (subwavelength gratings). Both gratings were made by Norcada~\cite{Norcada} on a low-tensile stress, $156\pm 1$ nm-thick, 1 mm-square SiN film suspended on a 500 $\mu$m-thick, 5 mm-square silicon frame. The gratings are 400 $\mu$m$\times$400 $\mu$m and possess trapezoidal fingers with period $850\pm 2$ nm, top and bottom finger widths $622\pm 6 $ and $678\pm 7$ nm and depth $46\pm 2$ nm, as determined by AFM profilometry~\cite{Darki2021} on a sample from the same fabrication batch. The refractive index of the SiN film, determined via ellipsometry, is $2.14\pm 0.01$ in the relevant wavelength range.

The best fit parameter results of the measured transmission and reflectivity spectra of both gratings, shown in Fig.~\ref{fig:gratings}, with the model of Sec.~\ref{sec:theory} are reported in Table~\ref{tab:gratings}. The minimum transmission and maximum reflectivity levels are 5.3\% and 91.6\% for grating A and 5.2\% and 90.9\% for grating B, respectively. Both gratings have thus very similar parameters in terms of losses and Fano resonance linewidth, and the 0.1 nm difference in the location of their Fano resonances show that they are spectrally overlapping.

\begin{table}
\centering
\caption{Best fit parameter results of the spectra of Fig.~\ref{fig:gratings} with the model of Sec.~\ref{sec:grating_model}.}
  \label{tab:gratings}
  \begin{tabular}{|c|cccccc|}
   \hline
   Grating & $t_d$ & $\lambda_0$ (nm) & $\lambda_1$ (nm) & $\gamma_\lambda$ (nm) &  $\beta$ ($\mu$m$^{-1}$) & $L$\\\hline
   $A$ & 0.813(3) & 951.855(1) & 951.997(1) & 0.482(1) & $9.6(1)\times 10^{-7}$ & 5.1(9)\% \\
   $B$ & 0.814(7) & 951.764(2) & 951.901(2) & 0.462(3) & $8.9(2)\times 10^{-7}$ & 4.1(8)\% \\\hline
  \end{tabular}
\end{table}

\subsection{Fano cavity setup}

\begin{figure}[h]
\centering
\includegraphics[width=0.6\columnwidth]{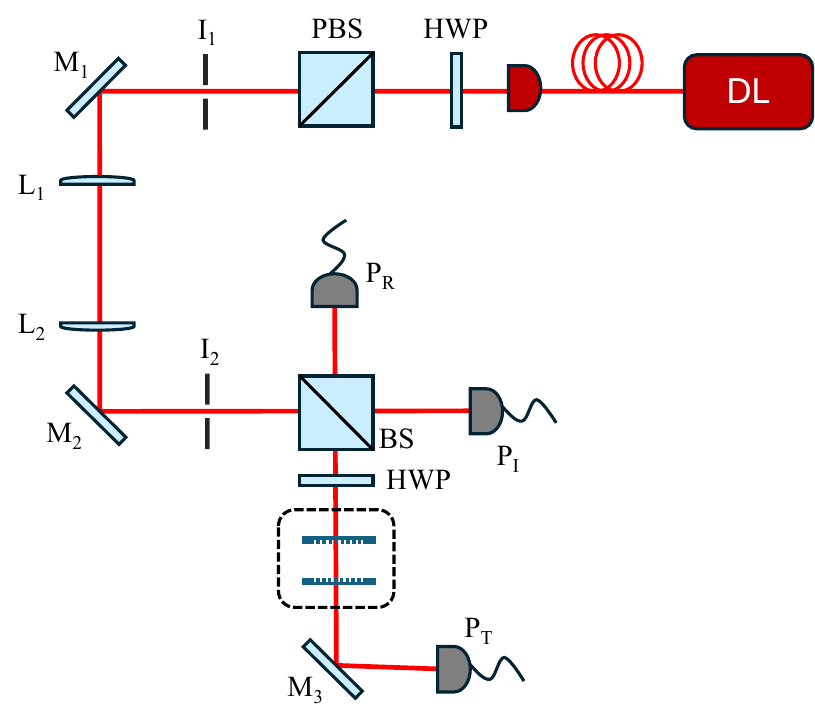}
\caption{Experimental optical layout. DL: diode laser. HWP: half-wave plate. PBS: polarizing beamsplitter. M$_1$, M$_2$, M$_3$: mirrors. L$_1$, L$_2$: lenses. BS: 50:50 beamsplitter. P$_\textrm{R}$, P$_\textrm{I}$, P$_\textrm{T}$: photodetectors.}
\label{fig:optical_layout}
\end{figure}

\begin{figure}[h]
\centering
\includegraphics[width=\columnwidth]{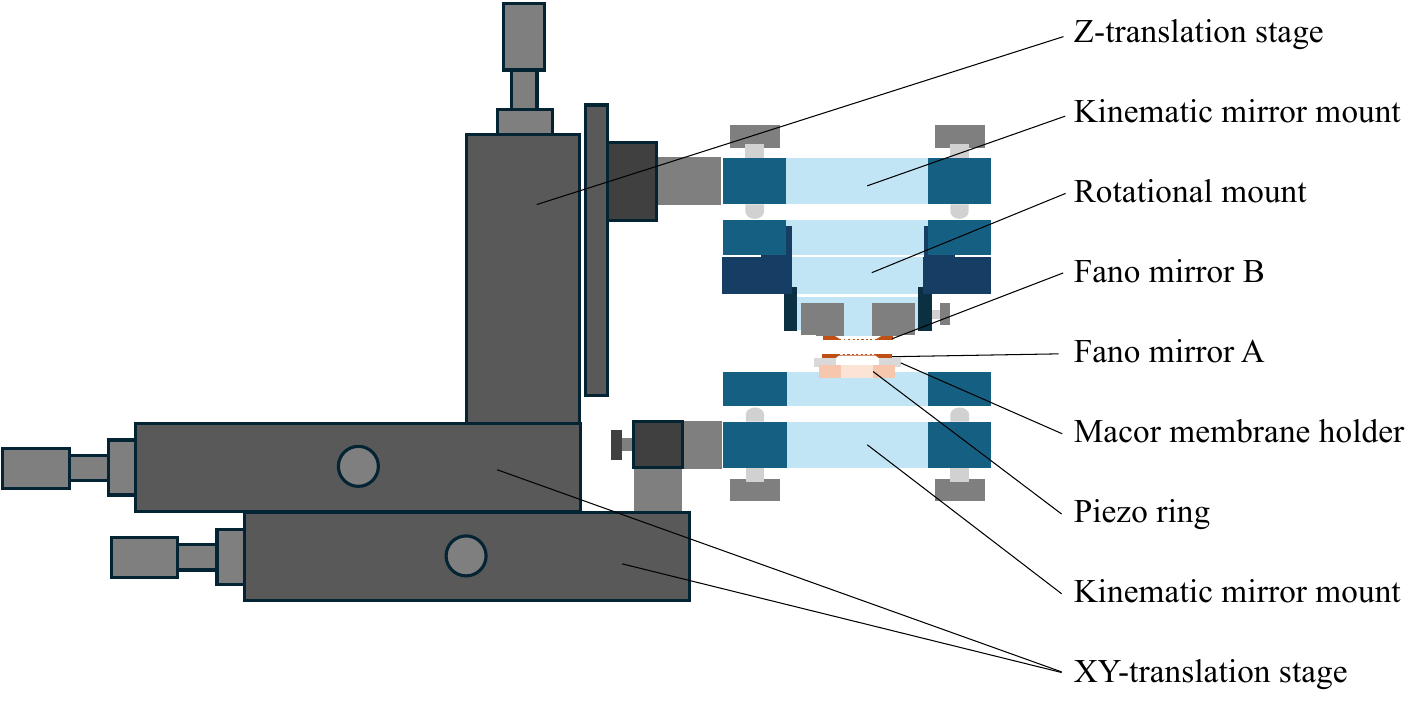}
\caption{Schematic of the cavity setup.}
\label{fig:cavity_setup}
\end{figure}

The grating/cavity transmission measurements were performed using the setup shown in Figs.~\ref{fig:optical_layout} and \ref{fig:cavity_setup}, which is an extension of the setup used in Ref.~\cite{Mitra2024} to realize single Fano mirror cavities. 

After spatial filtering in a single mode fiber, linearly polarized monochromatic light from a tunable diode laser (Toptica DLC CTL 900) is weakly focused (beam diameter $160\pm 10$ $\mu$m) on the grating/cavity (Fig.~\ref{fig:optical_layout}). The light transmitted and reflected by the grating/cavity is detected by photodiodes $P_t$ and $P_r$, respectively, which are referenced to the input power $P_i$ measured in transmission of the incoupling beamsplitter (BS).

As shown in Fig.~\ref{fig:cavity_setup}, the cavity setup consists of three adjustment mounts; two kinematic mirror mounts (Thorlabs Polaris-K13SP) are used to align the two reflective elements (grating/top mirror or HR/top mirrors), and a linear translation stage (Thorlabs NFL5DP20) is used to vary the distance between them. The linear stage is moved by a differential adjuster with a resolution of 50 $\mu$m per revolution that allows for varying the cavity length smoothly so as to be able to count the individual interference fringes. The mounts are also equipped by piezoactuators on top of their adjusters allowing for controlling their position with sub-micron resolution. The whole setup is installed on two XYZ translation platforms that are used to center the gratings with respect to the incident beam. A rotational mount is also installed on the top mirror mount in order to control the alignment of the grating fingers with respect to the incoming linearly polarized light. 

\subsection{Alignment procedure}

Double Fano mirror cavities are substantially more challenging to align than similar single Fano mirror cavities, such as the one realized in~\cite{Mitra2024}, due to the additional degrees of freedom introduced by the presence of the second Fano mirror. The procedure used to align the double Fano mirror cavities in this work is as follows:

\begin{itemize}
\item First, the bottom mirror $A$ is positioned and aligned using the bottom XY-translation stage and bottom kinematic mirror mount tilts, such that the incident beam is centered and normally incident on the grating, and the polarization of the incident light is set using the half-wave plate by minimizing the transmission level at the resonance wavelength of mirror $A$.

\item After marking the relative positions of the macor mount supporting mirror $A$ and the PZT mount, the macor mount with mirror $A$ is removed to make room for the alignment of mirror $B$ using transmission measurement.

\item The top mirror mount with mirror $B$ is inserted and aligned with respect to the fixed beam using the top XY-translation stage and top kinematic mirror mount tilts. In addition, the rotation mount is used to minimize the transmission at the resonance wavelength of mirror $B$, so as to ensure that the polarization and the grating fingers of mirror $B$ are correcty aligned with respect to each other.

\item The top mirror mount is removed, the bottom mirror mount reinserted and its alignement finely adjusted. The top mirror mount is then resinserted, the transverse and rotational degrees of freedom having been determined in the previous step. The tilt degrees of freedom are finely adjusted by aligning the back reflection from mirror $B$ with reference apertures I$_1$ and I$_2$ in the incident beam path.

\item The cavity length is adjusted using the Z-translation stage. A piezo ring actuator (Noliac NAC2124) below the bottom mirror is used in the alignment phase to scan the cavity length and adjust its value so that the cavity resonates approximately at the average of the Fano mirror resonant wavelengths. 

\item Once the distance between the cavity mirrors is fixed, broad and narrow wavelength scans are performed and the cavity transmission recorded. The broad-range scans allow for measuring the cavity Free Spectral Range (FSR) and calculating the cavity length, taking into account the wavelength dependence of the mirrors. The narrow-range scans allow for determining the cavity linewidth at the set cavity length. For cavity lengths less than $\sim 30$ $\mu$m, for which only one cavity resonance can be captured in the available laser wavelength range, the cavity length is inferred by counting the number of cavity resonance fringes observed when reducing the cavity length from the last length determined using the FSR method.
\end{itemize}

\subsection{Experimental results}

\begin{figure}
\centering
\includegraphics[width=0.6\textwidth]{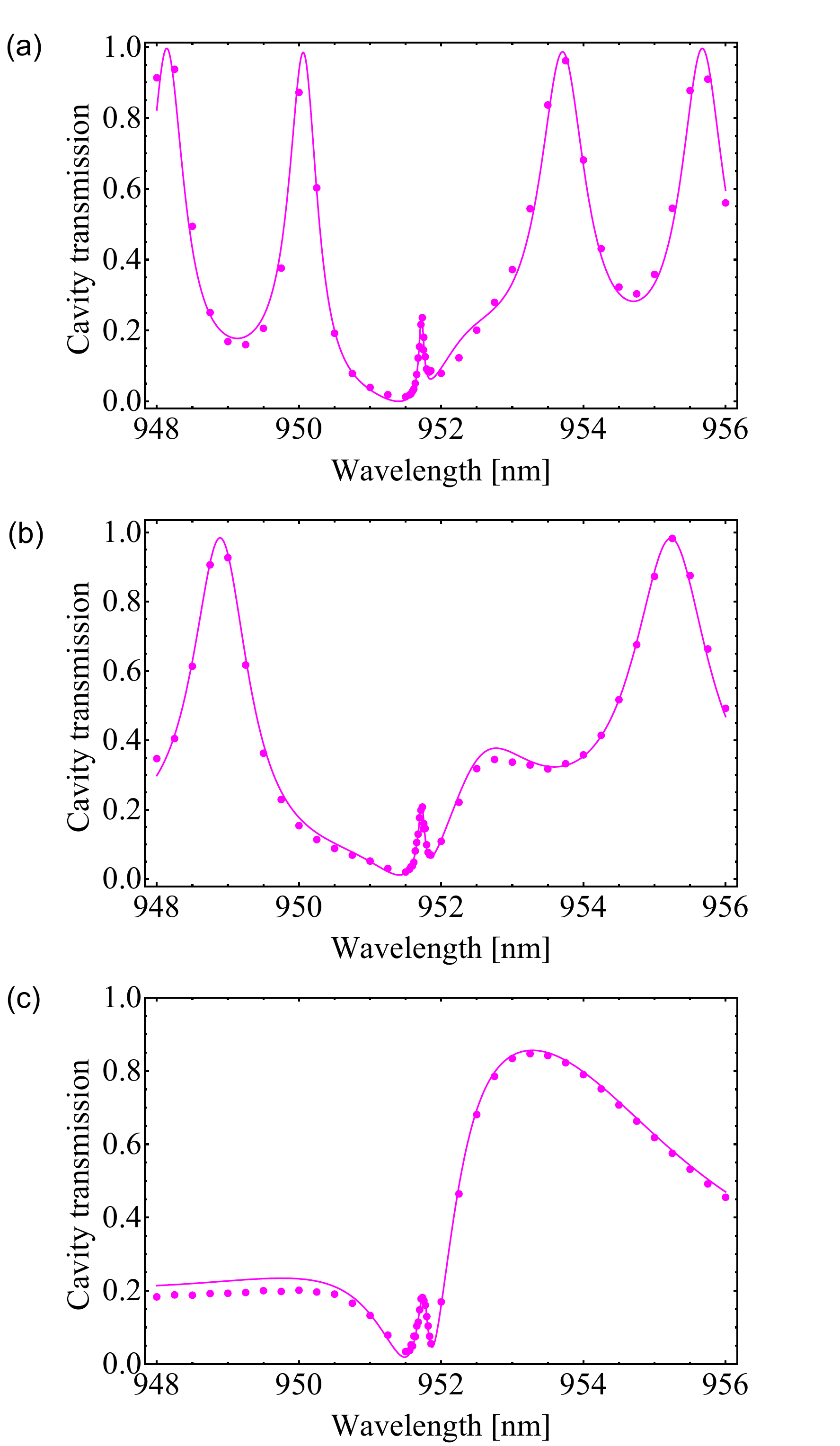}
\caption{Double Fano cavity transmission spectra for different cavity lengths: (a) $l=238.0$ $\mu$m, (b) $l=140.4$ $\mu$m and (c) $l=33.4$ $\mu$m. The lines show the result of a fit with the theoretical model (see text for details).}
\label{fig:double_spectra_long_exp}
\end{figure}

Figure~\ref{fig:double_spectra_long_exp} shows the normalized transmission spectra of a double Fano cavity consisting of Fano mirrors $A$ and $B$ for three different cavity lengths, $l=238.0$ $\mu$m, $l=140.4$ $\mu$m and $l=33.4$ $\mu$m. For the two longer cavity lengths the low-finesse Fabry-Perot fringes away from the Fano resonance are clearly visible in the accessible wavelength range. Close to the Fano resonance a much sharper peak emerges as a result from the coupling of the Fabry-Perot mode with the guided modes of the gratings, as was discussed in Sec.~\ref{sec:single_double}. The lines show the results of a fit with Eq.~(\ref{eq:Tcav}) using the grating model parameters of table~\ref{tab:gratings} and leaving the cavity length $l$ as a free parameter. In order to account for possible small deviation from normal incidence and increased losses due to misalignment or imperfect parallelism, the $\beta$ parameters are left free and a free offset is added to the resonant wavelengths.

There is an overall very good agreement between the measured spectra and the theoretical predictions. Among the possible reasons to explain the small discrepancies are {\it (i)} transverse misalignement between the gratings and the beam, {\it (ii)} deviation from normal incidence illumination, {\it (iii)} imperfect polarization matching, {\it (iv)} imperfect parallelism between the Fano mirrors, {\it (v)} collimation and finite-size effects or {\it (vi)} mechanical vibrations during the scans. We estimate that the effect of {\it (i)} and {\it (iii)} should typically be negligible (extra losses less than 1\%), given the translation stage and rotation mount accuracy. We expect the effect of {\it (ii)} to be potentially more pronounced, as the final alignment of the top mirror using the back-reflection is not as precise as the minimization of the grating transmission. Such a deviation from normal incidence would cause a shift of the Fano resonance and an increase in the losses at $\lambda_c$, potentially by a few percent. To minimize the effect of {\it (iv)} the tilts of both mirrors are finely adjusted around their initial alignment positions using rapid cavity linewidth measurements when the cavity length is scanned using the PZT transducer. The effects of {\it (iv)} on the cavity linewidth, as discussed in~\cite{Naesby2018}, are expected to be negligible here (extra losses less than 1\%), given the large waists and grating sizes used (see also~\cite{Mitra2024}). Regarding {\it (vi)}, to reduce the effect of mechanical vibrations, the whole setup is placed in a sealed glovebox and low-noise translation stages are used; however, as will be discussed shortly, the complexity of the double Fano cavity setup, which involves a large number of degrees of freedom, makes it still substantially more sensitive to mechanical noise.

\begin{figure}
\centering
\includegraphics[width=0.6\textwidth]{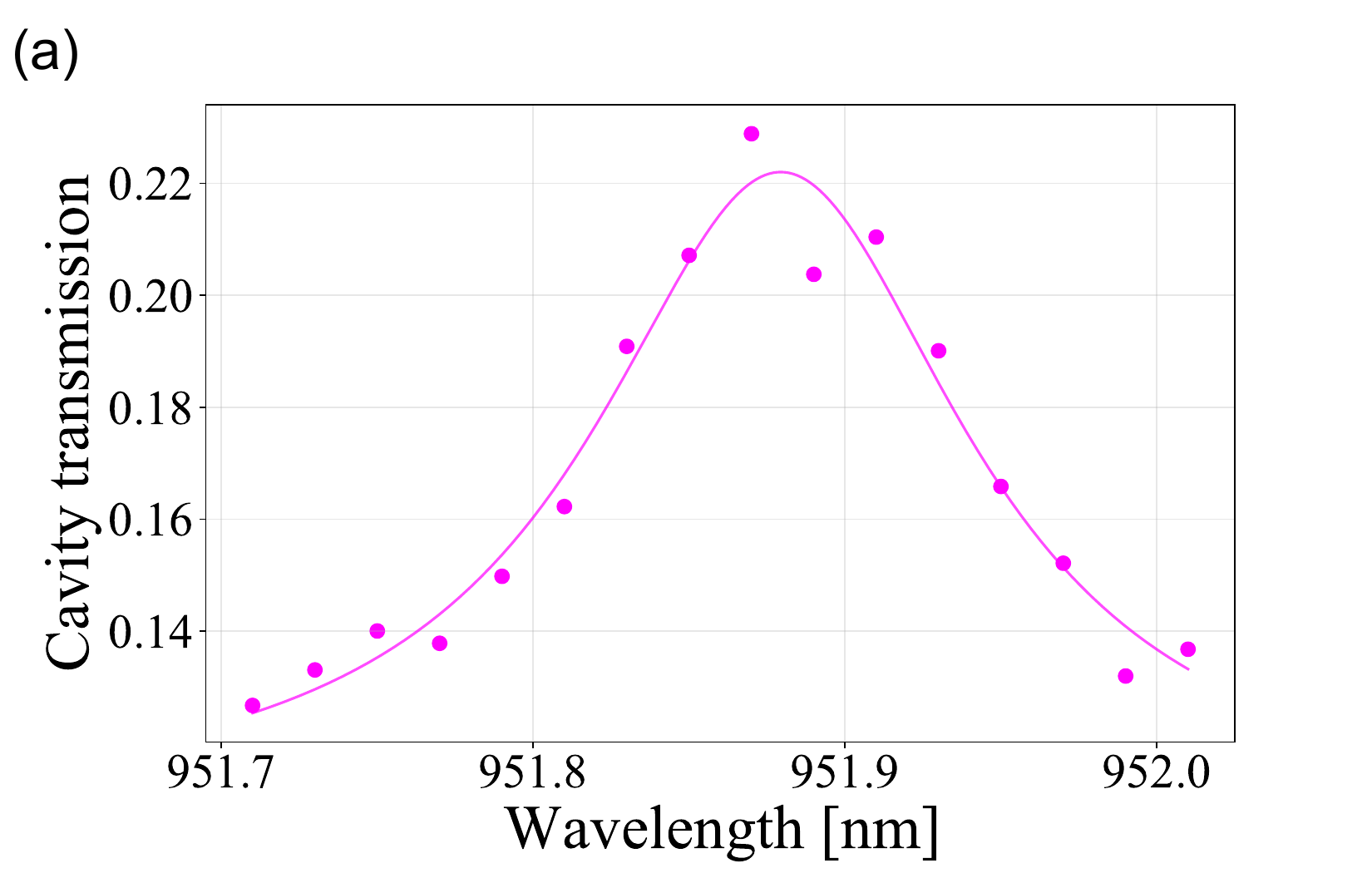}\\
\includegraphics[width=0.6\textwidth]{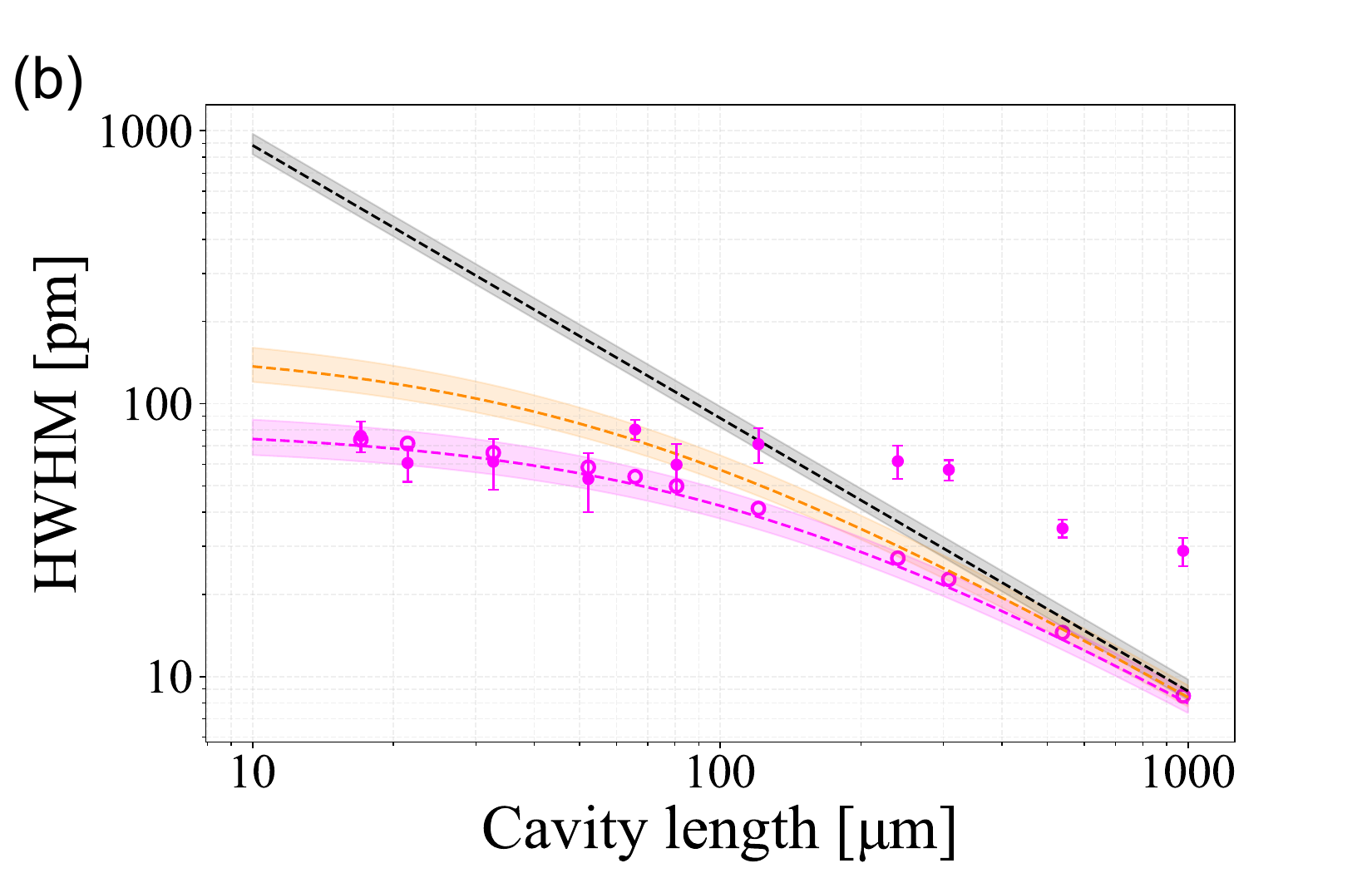}
\caption{(a) Example of a transmission spectrum around resonance for a cavity length of $17.0\pm0.2$ $\mu$m. The result of a fit with Eq.~\ref{eq:cavitylinewidth} yields $\delta\lambda=73\pm 10$ pm. (b) Experimentally measured cavity linewidth (HWHM) as a function of cavity length (magenta dots). Each point is the average of the results of 5-8 short range scans around resonance at a given cavity length. The magenta dashed line and the magenta empty circles respectively show the analytical and simulated predictions based on the measured grating parameters and cavity lengths. For the sake of comparison, the orange and black dashed lines respectively show the linewidths of the corresponding single Fano mirror and broadband mirror cavities with the same total losses.}
\label{fig:double_linewidth_exp}
\end{figure}

Narrow wavelength scans around the average Fano resonance wavelength $\lambda_c=(\lambda_0+\lambda_0')/2$ were performed for different cavity lengths between 17 $\mu$m and 1 mm. An example of such a scan is given in Fig.~\ref{fig:double_linewidth_exp}(a). The average linewidths extracted from fits with Eq.~(\ref{eq:cavitylinewidth}) of typically 5-8 of these scans are shown in Fig.~\ref{fig:double_linewidth_exp}(b), together with the theoretical predictions using the full numerical simulations (magenta points) and the analytical predictions of Eq.~(\ref{eq:deltalambda2}) using the reflectivities at $\lambda_c$ of both Fano mirrors, i.e., $\mathcal{L}=2-R_g^A(\lambda_c)-R_g^B(\lambda_c)$. While the measured linewidths are in good agreement for the shortest cavity lengths a substantial broadening is observed at longer cavity lengths. We attribute this to the previously mentioned mechanical noise affecting the cavity transmission measurements during the wavelength scans. Such a noise is not included in the theoretical model and was also observed previously in our realization of single Fano mirror cavities at long cavity lengths, albeit with a much reduced amplitude due to the significantly less complex cavity setup. In principle, such the effect of this noise during the measurement could be reduced by an improved vibration isolation or by actively stabilizing the cavity.

For the sake of comparison the expected linewidths for the corresponding broadband mirror and single Fano mirror cavities possessing the same amount of losses are also shown in Fig.~\ref{fig:double_linewidth_exp}, which shows that a reduction in spectral selectivity can be achieved with a double Fano mirror cavity in the short cavity length regime, which is the regime of interest for Fano cavity applications. 

Let us stress again that, while the linewidth reduction with respect to the corresponding broadband mirror cavity can be substantial (a reduction factor of $\sim 15$ was demonstrated in~\cite{Mitra2024}), the relative reduction between a single and a double Fano mirror cavity is at most a factor of 2, as discussed in Sec.~\ref{sec:theory}. Narrower linewidths could be achieved with higher Q and/or higher reflectivity Fano mirrors, using dual-period grating structures as suggested in~\cite{Singh2024} for instance. Using Fano mirrors with lower losses would also improve the Fano resonance extinction ratio, which could be beneficial for sensing and optomechanics applications. We note here that, while the Fano mirrors used in this particular realization have relatively high losses (4-5\%), lower loss levels (<1\%) have been achieved with similar 1D photonic crystal mirrors~\cite{ToftVandborg2021}, albeit with lower Q resonances.

%%%%%%%%%%%%%%%%%%%%%%%%%%%%%%%%%%%%%%%%%%%%%%%%%%%%

\section{Conclusion}
\label{sec:conclusion}

We performed a theoretical investigation of the transmission properties of Fano microcavities made of a pair of spectrally degenerate resonant Fano mirrors possessing high-Q internal resonances and showed that significant linewidth reduction can be obtained at short cavity lengths as compared to corresponding cavities using broadband reflectivity mirrors or a single Fano mirror and a broadband mirror. We then reported on an experimental realization of such cavities using as Fano mirrors ultrathin, suspended SiN films patterned with a subwavelength grating structure. The observed cavity transmission spectra over a wide range of cavity lengths were found to be in good agreement with the theoretical predictions and the expected linewidth reduction was observed. In the future, higher Q cavities could be obtained by using even higher-Q Fano mirrors~\cite{Singh2024}, by in situ tuning of their spectral resonances~\cite{Nair2019} or by exploiting bound state in the continuum resonances~\cite{Peralle2024}.

Such Fano microcavities with low modevolume and high-spectral sensitivity may be attractive for sensing applications, e.g. gas pressure sensing~\cite{Naserbakht2019,AlSumaidae2021,Hornig2022,Salimi2024}, but also for cavity optomechanics with multiple membrane resonators~\cite{Xuereb2012,Xuereb2014,Nair2017,Gartner2018,Piergentili2018,Wei2019,Manjeshwar2020,Yang2020b,Sheng2021,Cao2025,Yao2025} or with resonant, suspended membranes, as recently proposed in~\cite{Cernotik2019,Fitzgerald2021} and currently investigated in various experiments~\cite{Sang2022,Xu2022,Zhou2023,Enzian2023,Manjeshwar2023,Singh2025}. Let us note, in this respect, that the grating membrane resonators used in these experiments exhibit high mechanical Q-factor ($>10^6$) modes in the 200 kHz-1 MHz range and that strong optomechanical interactions with such mechanical modes have recently been observed in single Fano mirror cavities~\cite{Manjeshwar2023,Singh2025}.

%%%%%%%%%%%%%%%%%%%%%%%%%%%%%%%%%%%%%%%%%%%%%%%%%%%%%

%\begin{backmatter}
%\bmsection{Funding}
\section*{Funding}
Novo Nordisk Fonden (grant NNF22OC0075937).

%\bmsection{Acknowledgments}
\section*{Acknowledgments}
We are grateful to Norcada for their assistance with the design and fabrication of the samples used in this study.

%\bmsection{Disclosures}
\section*{Disclosures}
The authors declare no conflicts of interest.

%\bmsection{Data Availability Statement}
\section*{Data Availability Statement}
Data underlying the results presented in this paper are not publicly available at this time but may be obtained from the authors upon reasonable request.

%\bmsection{Supplemental document}
%\section*{Supplemental document}
%See Supplement 1 for supporting content. 

%\end{backmatter}

\end{document}